\documentclass[pra,showpacs,twocolumn]{revtex4-1}
\usepackage{slashed}
\usepackage{mathrsfs}
\usepackage{amsfonts}
\usepackage{amsmath}
\usepackage{amssymb}
\usepackage{revsymb}
\usepackage{graphicx}
\usepackage{mathrsfs}
\usepackage{bm}
\usepackage{psfrag}
\usepackage{color}
\usepackage{hyperref}
\usepackage{natbib}

\usepackage[usenames,dvipsnames]{xcolor}

\newcommand{\be}{\begin{equation}}
\newcommand{\ee}{\end{equation}}
\newcommand{\ba}{\begin{array}}
\newcommand{\ea}{\end{array}}
\newcommand{\bea}{\begin{eqnarray}} 
\newcommand{\eea}{\end{eqnarray}} 
\newcommand{\bd}{\begin{displaymath}}
\newcommand{\ed}{\end{displaymath}}
\newcommand{\eps}{\varepsilon}

\newcommand{\trm}[1]{\textrm{#1}}

\newcommand{\av}[1]{\langle #1 \rangle}
\newcommand{\figref}[1]{Fig. \ref{#1}}

\newcommand{\eqnref}[1]{Eq. (\ref{#1})}

\newcommand{\ccfl}{\tsf{\tiny{CCF}}}
\newcommand{\monol}{\tsf{\tiny{MONO}}}
\newcommand{\Ecr}{E_{\trm{cr}}}

\newcommand{\tsf}[1]{\textsf{#1}}

\newcommand{\vkap}{\varkappa}
\newcommand{\vphi}{\varphi}

\newcommand{\Ai}{\trm{Ai}}
\newcommand{\Gi}{\trm{Gi}}
\newcommand{\sech}{\trm{sech}}

\newcommand{\phpol}{e}
\newcommand{\vthe}{\vartheta}
\newcommand{\antares}{\tsf{\small{ANTARES}}~}
\newcommand{\optional}[1]{}

\newcommand*\xbar[1]{%
  \hbox{%
    \vbox{%
      \hrule height 0.5pt % The actual bar
      \kern0.2ex%         % Distance between bar and symbol
      \hbox{%
        \kern-0.15em%      % Shortening on the left side
        \ensuremath{#1}%
        \kern-0.15em%      % Shortening on the right side
      }%
    }%
  }%
}

\begin{document}
\title{Vacuum birefringence in high-energy laser-electron collisions}
 
\author{B.~King}
\affiliation{Centre for Mathematical Sciences, Plymouth University, Plymouth, PL4 8AA, United 
Kingdom}
\email{b.king@plymouth.ac.uk}

\author{N.~Elkina}
\affiliation{Arnold Sommerfeld Center for Theoretical Physics, 
Ludwig-Maximilians-Universit\"at M\"unchen, Theresienstra\ss e 37, 80333 M\"unchen, Germany}
\email{nina.elkina@physik.uni-muenchen.de}

\date{\today}
\begin{abstract}
  Real photon-photon scattering is a long-predicted phenomenon that is being searched for in 
  experiment in the form of a birefringent vacuum at optical and X-ray frequencies. We 
  present results of calculations and numerical simulations for a scenario to measure this effect 
  using multi-MeV photons generated in the collision of electrons with a laser pulse. We find that 
  the birefringence of the vacuum should be measurable using experimental parameters attainable 
  in the near future.
\end{abstract}
\maketitle

Shortly after the discovery of the positron \cite{anderson33}, several authors 
suggested the possibility that real photons could scatter off one another through interaction with 
virtual electron-positron pairs \cite{halpern34}. First calculated for 
low- \cite{euler35,weisskopf36,heisenberg36} and high-energy \cite{akhiezer36,akhiezer37} photons 
propagating in a constant electromagnetic background and later in plane-wave backgrounds 
\cite{narozhny69,baier75a}, recent advances in laser technology have generated much interest in 
discovering this effect in experiment \cite{bernard00,pvlas12,rizzo13,schlenvoigt15}. That photons 
polarised parallel and perpendicular to the background polarisation have a different probability to 
scatter, is often referred to as \emph{vacuum birefringence}. When the source of photons is a 
laser pulse, a signal of birefringence is predicted to be observable from the induced ellipticity 
in the pulse's field \cite{heinzl06,dipiazza06,king10b,dinu14a,dinu14b,homma15}, the scattered 
photons' 
angular distribution \cite{king10a,monden11,hatsagortsyan11,king12,karbstein15b} and parametric 
frequency shift 
\cite{lundstroem06,king12,gies14,boehl14,boehl15,gies16} (for a review, the reader is referred to 
\cite{marklund_review06,dipiazza12,king15a}).
\newline

Experimental probes of vacuum birefringence have focussed exclusively on colliding photons with
lab energies much less than the electron rest mass 
\cite{bernard00,pvlas12,rizzo13,schlenvoigt15}. As the ellipticity scales
linearly with photon flux, the high photon flux 
available for these parameters is beneficial for measurement. However, the cross-section for 
photon-photon scattering also scales with the sixth power of the centre-of-mass energy
\cite{landau4}. 
\newline

In the current letter we show that a different approach to measuring vacuum birefringence 
using multi-MeV rather than X-ray or optical photons yields a stronger signal of this long 
sought-after effect, at parameters that are achievable with today's experimental facilities. We 
substantiate our 
claim with analytical calculation and numerical simulation.
\newline

Birefringence of optical materials can be expressed by photons experiencing two different 
refractive indices, depending on how the photon's polarisation is aligned to the symmetry of the 
material's structure. When a real photon with wavevector $k$ and phase $\phi=k\cdot x$ propagates 
in a linearly-polarised plane wave background of vanishing frequency with wavevector $\vkap$, phase 
$\vphi=\vkap\cdot x$ and gauge 
potential $A=a_{1}g_{1}(\vphi)+a_{2} g_{2}(\vphi)$ where $\vkap\cdot a_{1,2} = 
a_{1}\cdot a_{2} = 0$, the two vacuum refractive indices experienced by a photon are for 
polarisation directions \cite{baier75a}:
\bea
\phpol_{1,2}^{\mu} = \frac{k \cdot \vkap ~a_{1,2}^{\mu} - k\cdot a_{1,2}\, 
\vkap^{\mu}}{k \cdot \vkap~\sqrt{-a_{1,2}^{2}}}.
\eea
($\hbar=c=1$ unless occurring explicitly.)
The locally-constant-field approximation (LCFA) of integrating the rate of constant-crossed-field 
(CCF)
processes over 
the spacetime structure of a non-constant laser background is believed to be valid 
\cite{ritus85,king13b,harvey15} when the classical intensity parameter $\xi = \sqrt{\alpha}
~|p\cdot F|/(m~p\cdot \vkap)$ \cite{ilderton09} ($\alpha\approx1/137$ is the fine-structure 
constant) fulfills $\xi\gg1$. 
Choosing $g_{2}(\vphi) = 0$, for a photon counterpropagating with the background, 
$\phpol_{1}$ ($\phpol_{2}$) is parallel (perpendicular) to the background electric field. The 
refractive index experienced by photons in these polarisation eigenstates is $n_{1,2}=1+\delta 
n_{1,2}$ where \cite{baier75a}:
\begin{align}
\begin{split}
\delta n_{1,2}(\vphi) = \frac{-\alpha m^{2}}{3(k^{0})^{2}} \int_{4}^{\infty} 
dv~\frac{z(\vphi)(2v+1\mp3)}{v\sqrt{v(v-4)}}f(z^{-1}(\vphi)), \label{eqn:dn12}
\end{split}
\end{align}
$z=(\chi_{k}/v)^{2/3}$, $\chi_{k}= \sqrt{\alpha}~|k\cdot F|/m^{3}$ is the quantum non-linearity 
parameter \cite{ritus85}, $F$ is the 
external-field Faraday tensor \cite{jackson75}, $m$ is the electron mass and $f(\cdot) = 
i\Ai(\cdot)+\Gi(\cdot)$ ($\Ai$ and $\Gi$ are Airy and Scorer functions of the first kind 
\cite{soares10}). When $\chi_{k}\ll1$, one finds from 
\eqnref{eqn:dn12} that $\delta n_{1,2}(\vphi) \approx \alpha (11\mp3) 
\chi_{k}^{2}/180\pi(k^{0})^{2}$, 
agreeing with well-known literature values \cite{baier67a}.
\newline

The polarisation $e$ of a real photon propagating through the birefringent vacuum can be 
expressed in terms of a superposition of two linear polarisation eigenstates \cite{landau4}:
\bea
e = 
\frac{1}{\sqrt{2Vk^{0}}}\left[\mbox{e}^{i\phi_{1}}\cos\vthe_{0}~\phpol_{1} + 
\mbox{e}^{i\phi_{2}}\sin\vthe_{0}~\phpol_{2}\right], \label{eqn:psi1}
\eea
where $V$ is volume, $\vthe_{0}$ is the initial polarisation angle of the photon with respect to 
$\phpol_{1}$ and $\phi_{1,2}$ is the phase acquired by each polarisation component:
\bea
\phi_{1,2} = \phi(k^{2}=0)-\frac{(k^{0})^{2}}{k\cdot\vkap}\int_{\vphi_{i}}^{\vphi_{f}}\!dy~\delta 
n_{1,2}(y),
\eea
when the photon travels between external-field phases $\vphi_{i}$ and $\vphi_{f}$.\optional{This 
can 
be written in a covariant way by defining four-vectors $\kappa_{1,2}$ in the lab rest frame as
$\kappa_{1,2}=(0,0,0,\delta n_{1,2})$, so that $\phi_{1,2}-\phi(k^{2}=0)=
\kappa_{1,2}\cdot x$.} As the photon 
propagates, a phase difference develops between the polarisation components, implying 
that an initially linearly-polarised photon becomes increasingly elliptically-polarised. However, as 
this change is just a pure phase, the mod-square and hence the probability that the 
photon polarisation is measured in a given eigenstate, remains constant as it must for an
eigenstate. Nevertheless, if the circular polarisation 
of a photon, which was initially linearly polarised at an angle $\vthe_{0}$ is measured after 
propagation in the birefringent vacuum, the probability $\tsf{P}_{\pm}$ of it being in a 
\emph{helicity} eigenstate $\phpol_{\pm} = (\phpol_{1}\pm i\phpol_{2})/\sqrt{2}$ is:
\bea
\tsf{P}_{\pm} = \frac{1}{2}\left[1\pm U(\vthe_{0},\Delta\phi)\right], 
\label{eqn:Ppm}
\eea
where $U(\vthe_{0},\Delta\phi) = \sin 2\vthe_{0}\sin\Delta\phi$ is the 
second Stokes parameter \cite{landau4} and $\Delta\phi=\phi_{2}-\phi_{1}$ is the phase lag induced 
by vacuum birefringence. Hence in a linearly-polarised background ``helicity flipping'' \emph{does} 
occur 
\cite{dinu14a, dinu14b}, unless the photon is prepared in a linear polarisation 
eigenstate. Measurement of a helicity flip in single photons should be contrasted 
with ongoing \cite{pvlas12,rizzo13} and planned \cite{schlenvoigt15} measurements of the 
ellipticity induced in the electromagnetic field of a laser wave, which is instead a collective 
effect on 
many photons. 
\newline

\begin{figure}[!h] 
\centering
\includegraphics[draft=false,width=8cm]{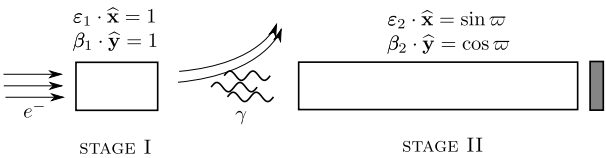}
 \caption{A schematic of the envisaged two-stage set-up to measure vacuum birefringence in 
high-energy photons. $a_{1}=(0,\pmb{\eps})$ and $a_{2}=(0,\pmb{\beta})$ are the polarisation 
directions of the external field in each stage.}
 \label{fig:setup1} 
\end{figure}

No helicity flip occurs for photons generated in a polarisation eigenstate. Therefore the
polarisation of the background used to generate photons must be different to the 
polarisation of the background in which the photons helicity-flip.\optional{As a consequence, 
helicity-flipping should not influence the development of an electromagnetic pair-cascade in 
a background with constant polarisation eigenstates.} To measure helicity-flipping 
in experiment, one could envisage a two-stage set-up in which the generation of high-energy 
photons and process of helicity-flipping are separated, such as depicted in \figref{fig:setup1}.
Stage I generates high-energy photons via nonlinear Compton scattering 
of electrons in an intense linearly-polarised counterpropagating laser pulse, and stage 
II collides the generated photons with a long counterpropagating laser pulse of linear polarisation
rotated by $\varpi$ in the plane of the first stage's polarisation. To demonstrate this set-up, we 
simulate both stages using the \antares particle-in-cell code that includes stochastic 
quantum effects using Monte Carlo methods, further details of which can be found in 
\cite{nerush11,elkina11}.
\newline

\begin{figure}[!h] 
\centering
\includegraphics[draft=false,width=8.6cm]{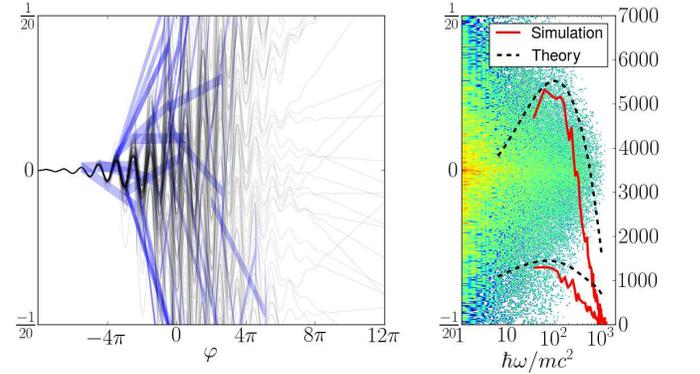}
 \caption{Left: electrons (black) and photons (blue) produced in stage I of the simulation with 
$\varkappa^{0} y/\pi$ as the vertical scale. 
Right: photon angular distribution (the left axis, with units $k_{y}/\pi$) and energy spectrum 
(the right axis, with units $\hbar \omega/mc^{2}$) of photons with polarisation $\phpol_{1}$ (upper 
two curves) and $\phpol_{2}$ (lower two curves).}
 \label{fig:trajel}
\end{figure}

\begin{figure}[!h] 
\centering
\includegraphics[draft=false,width=8cm]{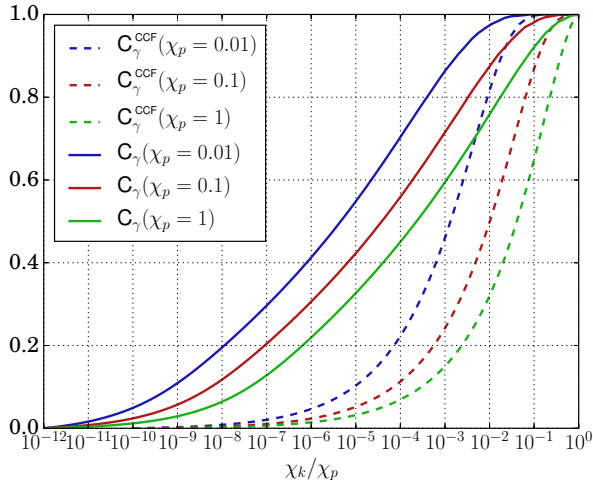}
 \caption{The normalised cumulative distribution of photons at the end of stage I (solid lines) and 
the corresponding distribution in a CCF (dashed line).}
 \label{fig:Cplot}
\end{figure}

By way of example for stage I, we simulate $2\,\trm{GeV}$ seed electrons 
counterpropagating with a laser pulse of frequency $1.55\,\trm{eV}$ ($800\,\trm{nm}$) and intensity 
parameter $\xi_{1}(\vphi)=\xi_{1}\,\exp[-(\vphi/\sigma_{1})^{2}]~|\!\cos\vphi|$ for $\xi_{1}=100$ 
where $\sigma_{1}=8\pi$ ($10.7\,\trm{fs}$), with $\pmb{\eps}_{1}=(1,0,0)$ and 
$\vkap_{1}=\vkap^{0}(1,0,0,1)$. The corresponding spectra and simulation dynamics are plotted in 
\figref{fig:trajel}. $2000$ unpolarised seed electrons generated 
a total of around $40000$ photons with polarisations parallel to the eigenstates 
$\phpol_{1}$ ($12130$ photons in the simulated spectrum in the plot, compared to 
$15740$ from theory) or $\phpol_{2}$ 
($3250$ photons in the simulated spectrum in the plot, compared to $4740$ from theory) via 
polarised nonlinear Compton 
scattering.  
The unpolarised photon spectrum expected from theory after stage I is:
\bea
\frac{\partial \tsf{P}_{\gamma}(\chi_{p},\chi_{k})}{\partial \chi_{k}} = 
\int_{-\infty}^{\infty} d\vphi~ 
\frac{\partial\tsf{R}^{\ccfl}_{\gamma}\left[\chi_{p}(\vphi),\chi_{k}
\right ] } {
\partial \chi_{k}}, \label{eqn:theoryspec}
\eea
where here $\chi_{p}=\chi_{p}(0)$ and $\tsf{R}^{\ccfl}_{\gamma}$ is the unpolarised rate per unit 
external-field phase for nonlinear Compton scattering in a constant crossed field. This is given by 
$\tsf{R}^{\ccfl}_{\gamma}(\chi_{p},\chi_{k}) 
= (\tsf{R}^{\ccfl}_{\gamma}(\chi_{p},\chi_{k}, 
\phpol_{1})+\tsf{R}^{\ccfl}_{\gamma}(\chi_{p},\chi_{k}, \phpol_{2}))/2$ for \cite{nikishov64}:
\bea
\frac{\partial\tsf{R}^{\ccfl}_{\gamma}(\chi_{p},\chi_{k},\phpol_{1,2}) } {
\partial \chi_{k}} &=& \frac{-\alpha}{\chi_{p}^{2}} 
\left\{\left[\frac{2\pm1}{z_{\gamma}}+\chi_{k}z_{\gamma}^{\frac{1}{2}}\right]\Ai'(z_{\gamma}
)\right.\nonumber\\
&& \left. \qquad\qquad\qquad\qquad +\Ai_{1}(z_{\gamma})\right\}
\eea
where $\Ai_{1}(x) = \int^{\infty}_{0} \Ai(t+x) dt$  and 
$\partial\tsf{R}^{\ccfl}_{\gamma}(\chi_{p},\chi_{k})/\partial \chi_{k}$ refers to the polarisation 
average of the rate. The normalised cumulative distribution after stage I:
\bea
\tsf{C}_{\gamma}(\chi_{p},\chi_{k}) =
\int_{0}^{\chi_{k}}\frac{\partial}{\partial \chi_{k}'} 
\frac{\tsf{P}_{\gamma}(\chi_{p},\chi_{k}')}{\tsf{P}_{\gamma}(\chi_{p})}~ d\chi_{k'}
\eea
is plotted with the normalised cumulative distribution for a CCF in \figref{fig:Cplot}. This 
highlights the much broader spectrum of frequencies produced in the oscillating pulse of stage I 
when the CCF is integrated over.
\newline

The proportion 
of photons in polarisation eigenstates $\phpol_{1,2}$ 
was determined by the ratio of the rates for nonlinear Compton scattering into those polarisations. 
The agreement with the spectrum produced in simulation is demonstrated by the dashed line 
in \figref{fig:trajel}. In the theoretical estimate, electrons are assumed to counterpropagate 
with the laser pulse. If the transverse excursion, which is included in simulation and shown in 
the figure, is included in the estimate, the electron quantum nonlinearity parameter $\chi_{p}$ 
should on average be reduced, leading to a lower estimate of the spectrum of photons produced.
\newline

\begin{figure}
     \includegraphics[width=4cm]{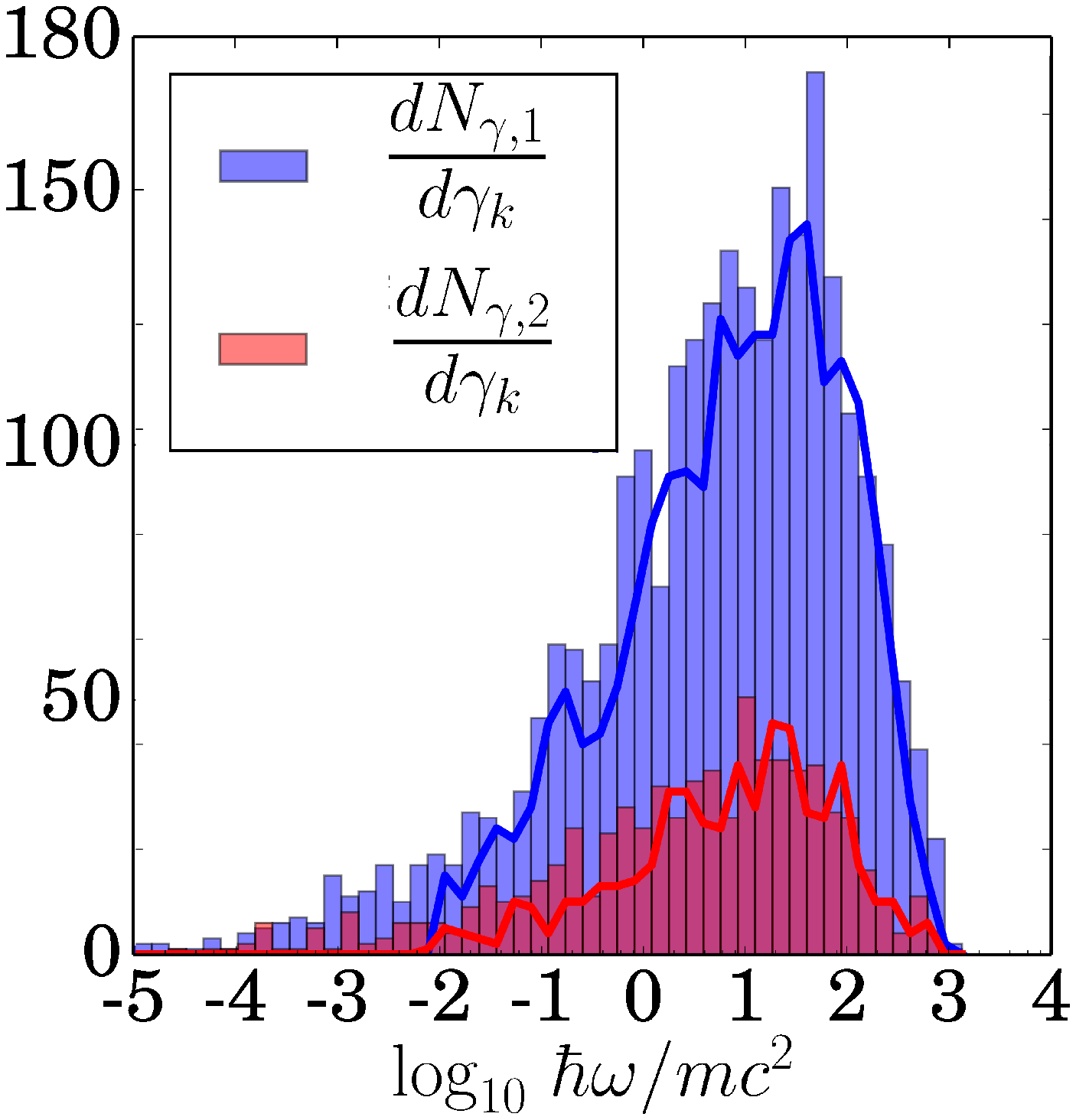}
   \includegraphics[width=4cm]{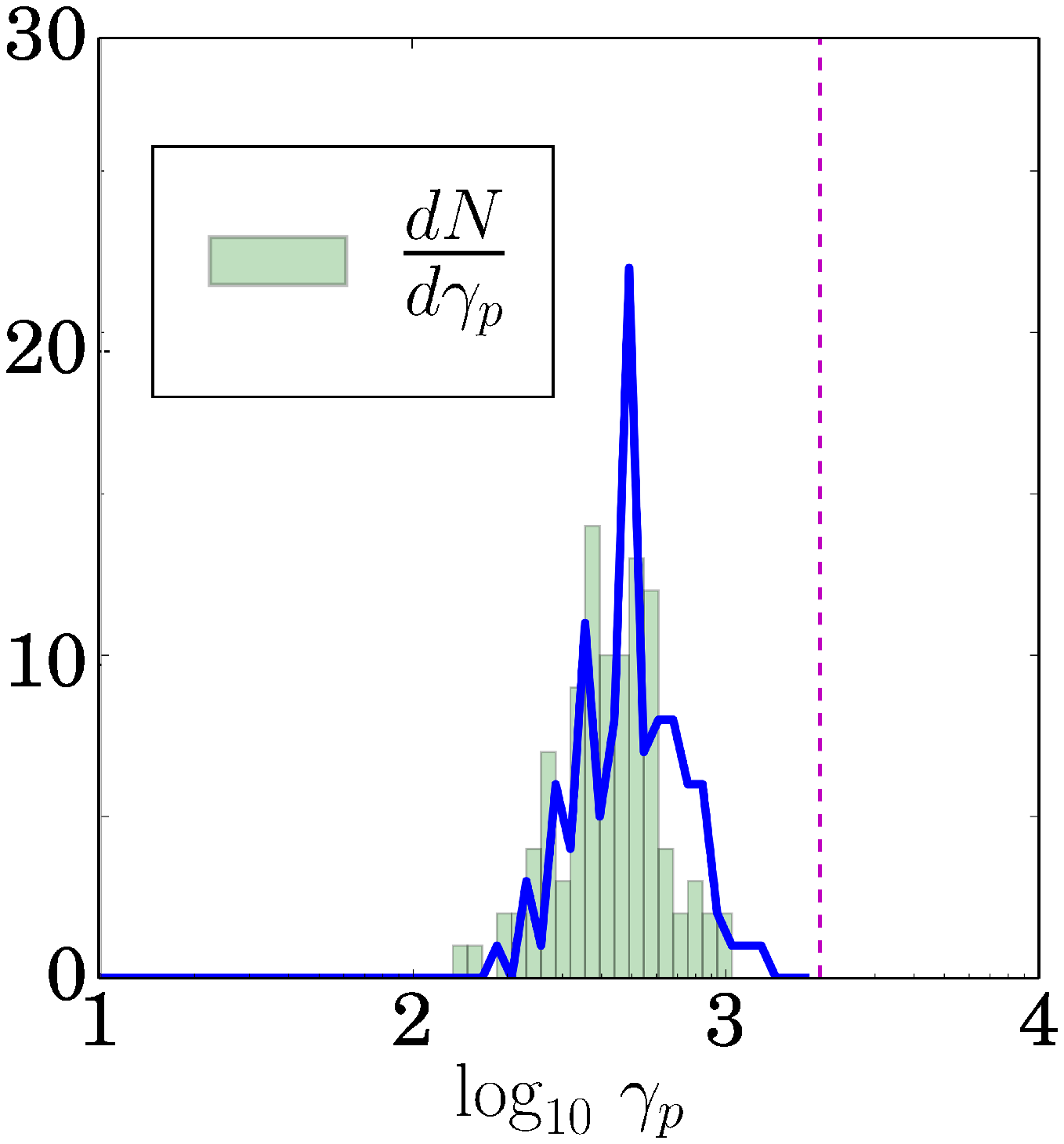}
  \caption{
Spectra obtained with the new event generator compared with stage I run for $100$ 
initial electrons of energy $500\,\trm{MeV}$, with $\xi_{1}=100$, $\sigma_1=8\pi$ and an 
initial electron offset of $z_0=\sigma_{1}$. $N_{\gamma,1}$ ($N_{\gamma,2}$) refer to photons 
created in the $\phpol_{1}$ ($\phpol_{2}$) eigenstate. Bars (lines) plot values using the new 
(old) event generator. $N_{\gamma,1}=2510$ ($2020$) and $N_{\gamma,2}=687$ ($513$) for the new 
(old) event generator.
    }
  \label{disc}
 \end{figure}

In order to describe the basic functionality of the simulation we 
follow the evolution of one event. In the event generator, we use three random 
numbers $0<r_1, r_2, r_3$. At the beginning of each time step
we first calculate the total unpolarised probability 
$\Delta 
t~\partial\tsf{P}^{\ccfl}_{\gamma}(\gamma_{p})/\partial t$ for an electron with gamma 
factor $\gamma_p$ to produce an unpolarised photon over time interval $\Delta t$. 
A new photon is created if this probability is greater than $r_{1}$. Numbers  $r_2$ and $r_3$ are 
then used 
to sample polarisation states and the energy of the secondary photon respectively. Which of the 
discrete 
polarisation states $\{\phpol_{1},\;\phpol_{2}\}$ that is assigned to a photon
is decided according to the relative expected abundance of each state:
$\tsf{P}^{\ccfl}_{\gamma}(\gamma_{p};\phpol_{1})/2\tsf{P}^{\ccfl}_{\gamma}(\gamma_{p})
\gtrless r_2$. The energy 
 of a polarised photon $k^{0}$, is sampled from the normalised cumulative distribution 
by solving the following equation 
 \begin{align}
\tsf{C}^{\ccfl}_{\gamma}(\gamma_{p},\gamma_{k};\phpol) = r_3.
\label{eqinversion}
\end{align}
Unlike our previous work \cite{elkina11} we do not apply any soft photon 
cutoff energy in calculations of singular integrals in corresponding probabilities. 
A new version of the event generator allows one to consider also soft photons by 
treating the weak singularity 
$\partial\tsf{P}_{\gamma}(\gamma_{p},\gamma_{k};\phpol)/\partial \gamma_{k} \sim \gamma_{k}^{-2/3}$ 
for $\gamma_{k} \ll \gamma_{p}$, by splitting the integration interval
into soft and hard parts. The soft part is evaluated using 
a change of integration variable from $\gamma_{k}$ to $\gamma_k^{1/3}$, which removes the apparent
singularity. Further improvements implemented in  accurate event generation routines will be 
described in 
\cite{elkina16}. Substantial improvement of the accuracy in QED simulation is a 
prerequisite for accurate predictions for proposals of future laser experiments.  To illustrate 
this point we apply our 
new  event generator for a collection of  photon data 
from stage I. In \figref{disc} we compare new results with ones obtained using the event 
generator 
of \cite{elkina11}.  
It can be seen that although qualitatively the results are similar, 
the quantitative difference in photon yield approaches 25\%. 
\newline

\begin{figure}[!h] 
\centering
\includegraphics[draft=false,width=8.6cm]{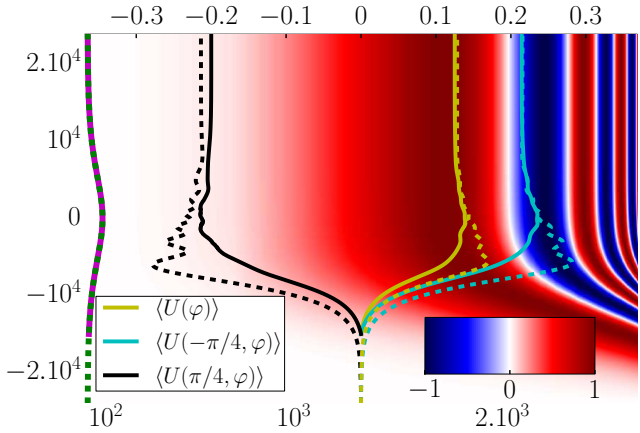}
 \caption{Simulation and theory results for stage II of \figref{fig:setup1}. At the end of stage 
II, simulation predicts $\av{U}=0.125$, theory predicts $\av{U}=0.128$. The vertical axis 
represents evolution in units of $\varphi/\pi$.  The background colour 
signifies $U$ for photon frequency given by the lower horizontal 
axis in units of the electron mass. The line along the left vertical axis is the shape of the laser 
background envelope. 
Dashed lines represent the calculations from theory, which assumes photons 
counter-propagate with the 
background, solid lines are results from simulation.}
 \label{fig:simres2} 
\end{figure}

\begin{figure}[!h] 
\centering
\includegraphics[draft=false,width=8cm]{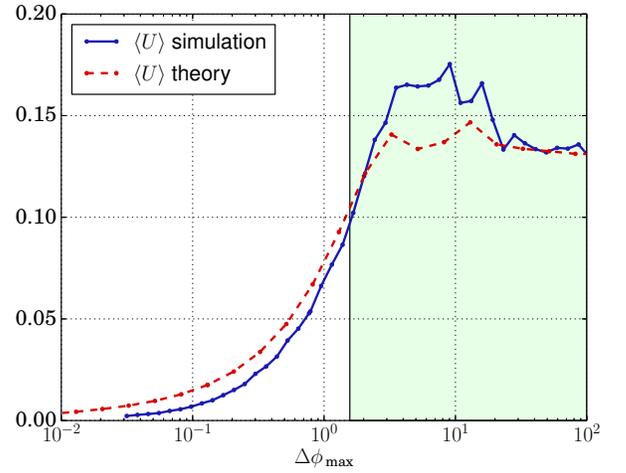}
 \caption{Average Stokes parameter after stage II. Theory predicts saturation of the effect to 
begin in the shaded region.}
 \label{fig:uplot1} 
\end{figure}

Following stage I, the electrons are filtered out and only the high-energy photons remain, which 
then collide with a long laser pulse. To illustrate the phenomena involved, we consider a laser 
frequency of
$1.55\,\trm{eV}$ 
($800\,\trm{nm}$) and 
classical nonlinearity parameter 
$\xi_{2}(\vphi)=\xi_{2}\,\sech^{2}(\vphi/\sigma_{2})~|\!\cos\vphi|$ for $\xi_{2}=50$, 
$\sigma_{2}=8000\pi$ ($10.7\,\trm{ps}$) and wavevector $\vkap_{2}=\vkap^{0}(1,0,0,1)$. Crucially, 
this laser pulse is now polarised with 
electric-field vector $\pmb{\eps}_{2} = (\sin\varpi, \cos\varpi,0)$ and we choose $\varpi=-\pi/4$ 
to 
maximise helicity flipping. To simplify the analysis, we make the approximation that photons 
collide head-on with the laser background in stage II, which will turn out to be a good 
approximation for the parameters we are considering. We consider the asymmetry 
$\tsf{P}_{+}-\tsf{P}_{-}$ to be the relevant 
experimental observable, which turns out to be exactly the second Stokes parameter $U(\Delta\phi)$ 
(since $\sin2\vartheta_{0}=-1$ for each photon in our discussion, from now on, we will suppress the 
$\vartheta_{0}$ 
argument in $U$). From \eqnref{eqn:Ppm} it is seen that this oscillates with propagation distance. 
Although photons in the $\phpol_{1}$ ($\phpol_{2}$) polarisation eigenstate will 
oscillate towards the negative (positive) helicity eigenstate $\phpol_{-}$ ($\phpol_{+}$) by equal 
amounts, since more $\phpol_{1}$ polarised photons are produced by stage I, the average ratio of 
each helicity eigenstate does not remain at $1/2$ and so $U$ does not remain at $0$. The 
evolution of the Stokes parameter through stage II, is plotted in 
\figref{fig:simres2}. As the 
photons begin to propagate through the background, the Stokes parameter increases
monotonically for all frequencies. However, $U$ oscillates over a shorter distance 
for higher frequency photons. This can be seen in by the more pronounced 
oscillation for higher photon frequencies in \figref{fig:simres2}. When the Stokes parameter of the 
highest-frequency photons changes 
sign, the spectrum average $\av{U}$ is no longer over terms of the same sign, the summation becomes 
incoherent and the effect saturates. No further increase in $\av{U}$ is expected beyond this point.
The average Stokes parameter measured after stage II is calculated from theory using:
\bea
\av{U}(\chi_{p}) = \tsf{A}_{\phpol}(\chi_{p})\int d \chi_{k} 
~U[\Delta\phi(\chi_{k})]\,\frac{\partial}{\partial\chi_{k}}\frac{\tsf{P}_{\gamma}
(\chi_ { p },\chi_ { k } ) }{\tsf{P}_{\gamma}(\chi_{p})}\,, \nonumber\\
\eea
for post stage II values, where $\tsf{A}_{\phpol}(\chi_{p}) = 
(\tsf{P}_{\gamma}(\chi_{p},\phpol_{1})-\tsf{P}_{\gamma}(\chi_{p},\phpol_{2}))/2\tsf{P}_{\gamma}
(\chi_{p})$ is the 
asymmetry in the photon polarisation. 
The agreement with simulation and the effect of saturation is shown in \figref{fig:uplot1}. For the 
relevant case of $\chi_{k}\ll1$, the phase difference has the 
simple form $\Delta\phi = \alpha \av{\chi_{k}\xi_{2}}_{\vphi}/30\pi$, where 
$\av{f}_{\vphi} = \int\,d\vphi~ f(\vphi)$. For the many-cycle pulse form of stage II, this 
becomes $\Delta\phi = \alpha\chi_{k}\xi_{2}\sigma_{2}/45\pi$. This implies in the lab system, the 
wavelength of oscillation of the Stokes' parameter $\lambda_{U}$ is related to the wavelength of 
the individual photon $\lambda$ by $\lambda/\lambda_{U} = 
(\alpha/30\pi)(E/\Ecr)^{2}(1-\cos\theta)$. It can be 
seen that several parameters of stage II can be combined into the single invariant parameter 
$\Delta\phi_{\trm{max}} = \alpha \av{\chi_{p}\xi_{2}}_{\vphi}/30\pi$, as plotted in 
\figref{fig:uplot1}. As $\av{U}$ is a ratio of probabilities generated by integration over the 
spectrum and the pulse, this is independent of $\xi_{1}$ in the CCF. For the LCFA to be applicable, 
the pulse used in stage I must be many-cycle and hence $\av{U}$ is also independent of $\sigma_{1}$ 
in the parameter range of interest. Remarkably, the single parameter $\Delta \phi_{\trm{max}}$ is 
sufficient for quantifying helicity-flipping for a large range of possible variables in the 
considered scenario. This allows one to be able to predict for what parameters $\av{U}$ saturates 
and for what value, depicted in \figref{fig:uplot1}.
\newline

\optional{In the measurement of scattering processes, it is common to consider emission directions 
where the 
signal is enhanced. On the one hand, an electron emits photons mostly in a cone of opening angle 
$\sim 1/\gamma_{p}$ around its direction of momentum \cite{jackson75}. On the other hand, the 
angle swept out by an electron's direction of momentum due to transverse 
excursion in the plane wave is of the order $\sim \xi/\gamma_{p}$. Since $\xi \gg 1$ for the LCFA 
to be valid, an approximation is made in simulation that photons are emitted parallel to the 
electron momentum. We 
investigated applying an angular cut to the photons measured at the end of stage II to enhance the 
signal, but found that although there is a change at very small angles, the change is both small 
and fluctuates between being positive and negative. We see in the maximum 
in the right-hand plot of \figref{fig:trajel} that although the highest-momentum photons, which 
have the highest Stokes parameter, are emitted in the centre of the angular distribution, this is 
also where the highest number of low-energy photons are emitted. For this reason we see no 
advantage in this scenario, to restricting the angular range of measured photons.}

\begin{figure}[!h] 
\centering
\includegraphics[draft=false,width=8cm]{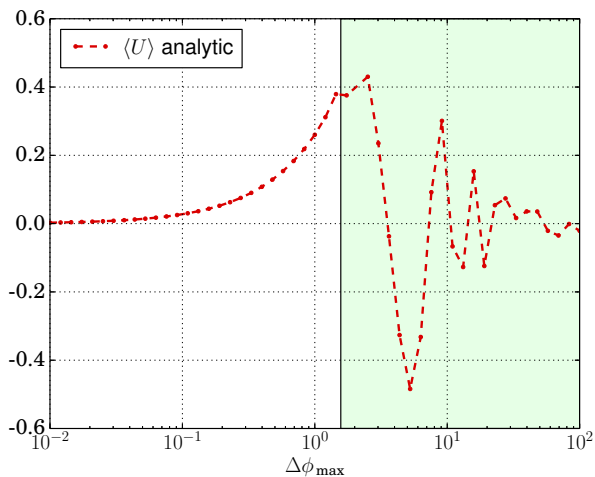}
 \caption{Average Stokes parameter after stage II for the accelerator set-up with $\xi_{1}=0.36$, 
and $\sigma_{1}=17\pi$ ($15\,\trm{fs}$). The shaded region corresponds to where theory 
predicts the onset of saturation.}
 \label{fig:Uplotmono1} 
\end{figure}

One way of increasing the signal and the maximum value of the Stokes parameter is to use a much 
narrower-bandwidth source of high-energy photons. An example of this is to take a
synchrotron as a source of high-energy photons, which has recently been analysed in 
\cite{ilderton16}. Another possibility is to consider the stage I electrons to be produced at 
higher 
energies in a particle accelerator. By way of example we briefly consider this
\emph{accelerator-based} set-up, taking values of the order of those achieved at 
the Stanford Linear Accelerator E-144 experiment \cite{burke97,bamber99} of
$47\,\trm{GeV}$ 
electrons colliding with a $1.6\,\trm{ps}$ laser pulse of maximum intensity 
parameter $\xi_{1}=0.36$ at $527\,\trm{nm}$ wavelength. Since $\xi_{1}\not\gg 1$ 
and the LCFA used in simulations is no longer 
valid \cite{dinu16}, we estimate the photon 
spectrum using \eqnref{eqn:theoryspec} with the unpolarised rate for nonlinear Compton 
scattering in a monochromatic background. The leading-order term in the perturbative 
expansion in $\xi$ is \cite{ritus85}:
\bea
\frac{\partial\tsf{R}^{\monol}_{\gamma}(\chi_{p},\chi_{k},\xi,\phpol_{1,2}) } {
\partial \chi_{k}} \approx
\frac{\alpha\xi^{2}}{16\pi\chi_{p}^{2}}\left[2\pm 1 +\frac{\chi_{k}^{2}}{\chi_{p}(\chi_{p}-\chi_{
k } ) }\right], \nonumber\\\label{eqn:dRmono}
\eea
where the polarisation dependency is similar to the CCF case, $\chi_{k} \in [0, 
\chi_{k}^{\trm{max}}]$ and $\chi_{k}^{\trm{max}}= 
\chi_{p}[1+\xi(1+\xi^{2}/2)/2\chi_{p}]^{-1}$, 
which is integrated over the pulse envelope in stage I\optional{It is 
straightforward to show \eqnref{eqn:dRmono} for a monochromatic wave is equivalent to taking the 
Klein-Nishina rate for linear Compton scattering of single photons and summing over the number of 
photons $E^{2}V/4\pi\vkap^{0}$ for electric field strength $E$ in a volume $V$}. For stage I, we 
consider a short pulse with $\xi_{1}=0.36$ and $15\,\trm{fs}$ duration and the same 
frequency as the E-144 experiment of $2.35\,\trm{eV}$ ($527\,\trm{nm}$ wavelength). Apart from 
being considerably 
suppressed, the produced spectrum differs from the LCFA case ($\xi \gg 1$) in that the 
mean 
frequency and variance are both lower. As the spectrum is narrower,
the value the Stokes parameter can reach before saturation is increased, as illustrated in 
\figref{fig:Uplotmono1}.
For example, if the laser employed in stage II was slightly longer than in E-144 at $10\,\trm{ps}$ 
($\sigma_{2}=11350\pi$), and slightly stronger at $\xi_{2}=1$ then $\Delta\phi_{\trm{max}}=0.57$ 
and $\av{U}=0.15$. This is comparable with the \emph{laser-based} set-up considered previously, if 
the same $\chi_{p}$ value is taken and $700\,\trm{MeV}$ seed electrons collide with a $\xi_{1}=50$, 
$\sigma_{1}=8\pi$, $1.55\,\trm{eV}$ pulse in stage I, followed by photons propagating through a  
$\xi_{2}=20$, $5.4\,\trm{ps}$ ($\sigma_{2}=4000\pi$) pulse in stage II then $\av{U}=0.10$. Although 
these values for the Stokes parameters are of similar magnitude, the accelerator-based set-up 
considerably relaxes the requirement on the stage II laser.
\newline

The helicity of photons with energies $1-10\,\trm{MeV}$ can be measured using Compton scattering on 
polarised atomic electrons in transmission polarimetry \cite{alexander09}. Furthermore, 
the creation of electron-positron pairs in an intense laser-pulse is also polarisation-dependent 
\cite{king13a} and this form of polarimetry in a similar type of set-up for parameters at the 
ELI facility has been considered recently in more detail in \cite{homma15}.
\newline

When high-energy photons propagate through an intense electromagnetic background, they can undergo 
decay into electron-positron pairs \cite{narozhny69}. These pairs would 
emit high energy photons and hence act as a background for the signal of photon-photon scattering. 
The expected number of pairs generated per electron in stage I via electron-seeded pair creation 
can 
be approximated for $\xi\gg1$ using $\av{N_{e}}=\sigma_{1} \tsf{R}_{\gamma e}(\chi_{p})$ with the 
approximate rate 
$\tsf{R}_{\gamma e}(\chi_{p})=3\alpha^{2}\ln(1+\chi_{p}/12)\exp(-16/3\chi_{p})(1+0.56\chi_{p}
+0.13\chi_{p}^{2})^{1/6} $ (adapted from \cite{baier72}), giving $\av{N_{e}}\approx 10^{-9}$ for 
the parameters considered in the laser-based set-up and is even lower for the accelerator-based 
set-up. For stage II, the probability per unit phase of pair creation 
for $\phpol_{1}$ and $\phpol_{2}$ polarised photons is $\tsf{R}_{e} \sim [\alpha \sqrt{3}(2\mp1)/8] 
\mbox{e}^{-8/3\chi_{k}}$ respectively \cite{king13a}, which is also heavily suppressed for 
$\chi_{k} \ll 1$. 
\newline

A further source of background is the nonlinear Compton scattering of any residual electrons in the 
synthetic vacuum of stage II. Although initially these electrons are effectively at rest and 
nonlinear Compton scattering is negligible since $\chi_{p} \ll 0.1$, they will be accelerated in 
the large field volume and could potentially mask the photon-photon scattering signal. For 
$\chi_{p} 
\ll 1$, the probability per unit phase of generating $\phpol_{1}$ and $\phpol_{2}$ polarised 
photons via nonlinear Compton scattering is \cite{king13a} $\tsf{R}_{e} = \alpha(5\pm 
3)/2\sqrt{3}$. For the long phase lengths considered in stage II, the expected number of photons 
generated per residual electron will be larger than one. However, since $\chi_{p}$ is in 
general lower than in stage I, these background photons will have a much lower frequency. Using an 
IR frequency cutoff on the photon-polarisation measured would remove this source of background. An 
example density of residual electrons is that in the LHC beamline, which is of the order of
$10^{6}\,\trm{cm}^{-3}$ \cite{brumfiel08}.
\newline

To conclude, the use of high-energy photons, such as produced in the collision of electron beams 
and laser pulses, offers a method significantly more sensitive to vacuum polarisation than current 
optical and forthcoming X-ray probes. As a consequence, such experiments can be used to place 
more stringent limits on the mass and interaction strength of axion-like particles, WISPs 
(Weakly Interactive Sub-eV Particles) and other dark-matter candidates 
\cite{villalba-chavez14,hu14b}. 
With the \antares code, we have demonstrated how 
real photon-photon scattering can be included in QED-plasma simulations and displayed good 
agreement with theory. Furthermore, we have showcased a new event generator for nonlinear Compton 
scattering that is capable of including arbitrarily low energies of photons.

B. K. would like to acknowledge the hospitality of N. Elkina and useful discussions with P. B\"ohl 
and A. Ilderton. This work was 
supported by the Grant No. SFB TR18 project B13 and the Arnold Sommerfeld Center for 
Theoretical Physics. Plots were generated with 
{\tt{Matplotlib}} \cite{matplotlib}.
%%%%%%%%%%%%%%%%%%%%%%

\optional{\section{Supplemental Material}
The gauge potential $e(k)$ of a photon with wavevector $k$ propagating in a constant crossed field 
satisfies the wave equation:
\bea
\square\, e_{\mu}(k) = \Pi^{\mu\nu}(k)e_{\nu}(k), \label{eqn:we1}
\eea
where $\Pi^{\mu\nu}$ is the elastic part of the polarisation operator and:
\bea
\Pi^{\mu\nu}e_{\nu} = 2(k^{0})^{2}\left(\delta n_{1} e_{1}^{\mu}e_{1}^{\nu} + \delta n_{2} 
e_{2}^{\mu}e_{2}^{\nu}\right) e_{\nu}.
\eea}
\optional{One can see that $\Pi^{\mu\nu}e_{1,2} = 2(k^{0})^{2} \delta n_{1,2} e_{1,2}$ so $e_{1,2}$ 
can 
be 
thought of as eigenpolarisations of the polarisation operator. It can also be shown that $e = 
e_{1,2}\mbox{e}^{i\phi_{1,2}}$ are independent solutions to \eqnref{eqn:we1} and therefore so 
is the linear combination chosen in the text.}

\bibliography{current}

\end{document}